# Investigation on the self-starting Mamyshev oscillator with low threshold

Wei Guo[1], Sihua Lu[1], Qiurun He[1], Baofu Zhang[2], Shanchao Ma[1], Zhongxing Jiao[1], and Tianshu Lai[1]

[1] School of Physics, Sun Yat-sen University, Guangzhou 510275, China
[2] Research Institute of Interdisciplinary Science & School of Materials Science and Engineering, Dongguan University of Technology, Dongguan 523808, China

E-mail: zhangbf@dgut.edu.cn, jiaozhx@mail.sysu.edu.cn

## Abstract

We have demonstrated a mode-locked Mamyshev oscillator based on a single-mode-fibre configuration with low threshold. The influences of spectral filters and the pump power on the output laser characteristics have been investigated. In the case with spectral-filter central wavelength far away from the peak of gain spectrum, the laser pulses have deviated slightly from similariton evolution with gain-shaping under high pump power. Moreover, self-starting mode-locking operation has been realized when the pump power was only 520 mW, which might benefit from the appropriate polarization conditions and small spectral filter separation. The laser can generate 1.89-nJ ultra-short pulses with 20-dB spectral width of 54.6 nm. The pulse duration can be compressed externally to 64.69 fs with the peak power of 21.3 kW.

## 1. Introduction

Benefit from compact structure, excellent heat dissipation and low-cost designs, ultrafast fibre lasers can produce high-peak-power, high-beam-quality, and ultra-short pulses. Therefore, they have played important roles in scientific research, ultra-precision manufacturing, and biomedical applications. Among them, ultra-short-pulse generation is commonly based on passively mode-locking mechanism by using material-based saturable absorbers [1, 2], nonlinear amplifying fibre loop mirror (NOLM) [3], and nonlinear polarization evolution (NPE) [4]. In addition, in order to achieve higher pulse energy and higher peak power in ultrafast fibre lasers, researchers have developed a variety of ultra-short pulse evolution forms, such as soliton [5], dispersion-managed soliton [6], self-similar [7], and dissipative soliton [8]. However, the accumulation of nonlinear effects in fibre will severely limit the output pulse energy and peak power of ultrafast fibre lasers mentioned above, and thus their performance still cannot challenge the commercial solid-state lasers [9].

Recently, researchers have turned their attention to a new type of ultrafast fibre laser, Mamyshev oscillator. Based on two offset spectral filters and self-phase modulation (SPM) [10, 11], Mamyshev oscillator can produce ultra-short pulses with greatly suppressing low-intensity pulses as well as continuous-wave lasing. Only laser pulses with sufficiently high peak power can generate enough spectrum broadening and pass through the offset filters. Mamyshev oscillators have been experimentally demonstrated to have great potential in generating high-energy ultra-short pulses. [12-23]

Nevertheless, the structure with two offset filters also brings the challenge of self-starting[12, 13]. In order to obtain self-starting in Mamyshev oscillators, researchers have proposed several methods including pump power modulation [14, 15], additional starting arm configuration[13, 16], external seed pules injection[17, 18]. However, most of these methods will compromise the output performances, and thus more attention should be paid on the mechanism and new methods of self-starting in Mamyshev oscillators. In addition, in order to improve output pulse energy, some research groups have investigated the influences of pump power on output performances[17-19], while others have focused on the parameters of filters implanted in the lasers [20, 21]. Using fibre Bragg gratings as spectral filters, Vincent Boulange et al.[20] and Željudevicius et al.[21] exploited the tunability of FBGs to achieve high-energy pulses. However, because of gain saturation, reabsorption, nonlinearity and their compound effects in the gain fibre, the influences of those cavity parameters on the pulse evolution of the Mamyshev oscillators remain unclear.

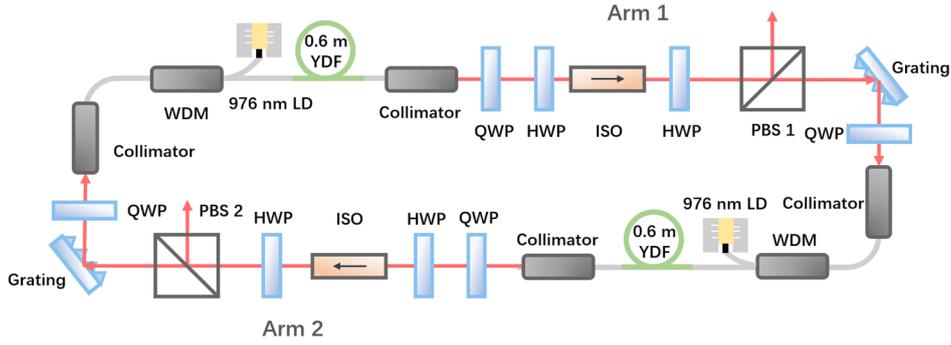

**Figure 1.** Schematic diagram of the Mamyshev oscillator. WDM: wavelength division multiplexer; LD: laser diode; YDF: ytterbium-doped fibre; QWP: quarter-wave plate; HWP: half-wave plate; ISO: isolator; PBS: polarizing beam splitter.

In this letter, we experimentally demonstrate an Yb-doped fibre Mamyshev oscillator with low threshold. We have investigated the influences of the spectral-filter central wavelength and the pump power on the output characteristics. Besides, the self-starting mode-locking operation could be achieved with total pump power of only 520 mW, which might benefit from the appropriate polarization conditions and small spectral filter separation in our laser. With the optimization, this Mamyshev oscillator could produce 1.89-nJ ultra-short pulses with a 20-dB bandwidth of 54.6 nm. The output pulses could be dechirped externally to 64.69 fs, and the signal to noise ratio of radio frequency (RF) spectrum was measured to be more than 70 dB.

## 2. Experimental setup

The Mamyshev oscillator presented here had a ring cavity which consisted of two Mamyshev regenerators (two arms) providing seed from one to another. Its configuration is illustrated in Figure 1. In each arm, a 976-nm single-mode laser diode with maximum output power of 400 mW was used as the pump. Its pump power was coupled into an Yb-doped fibre (YB1200-4/125) through a wavelength division multiplexer. The length of the Yb-doped fibre was designed to be 0.6 m due to a trade-off between high gain and severe reabsorption effects. The generating laser pulses would experience SPM-induced spectral broadening in the gain fibre and passive fibre, and then enter the free space section through a collimator. A quarter-wave plate (QWP) and a half-wave plate (HWP) were used to optimize the polarization of laser pulses for obtaining stable mode-locked operation. An isolator was placed after these two plates. It was used for not only obtaining the unidirectional operation of the laser, but also ensuring the NPE mechanism to achieve initiate mode-locked pulses. Moreover, the combination of another HWP and polarizing beam splitter (PBS) was taken as the output coupler of this arm. The output coupling ratio could be adjusted by rotating this HWP. The residual laser with broadened spectrum in the cavity would be reflected by a diffraction grating, and then pass through an input collimator into the next arm. This 300 lines/mm grating together with the input collimator could serve as a narrow Gaussian spectral band-pass filter, with the 3-dB bandwidth estimated to be ~3 nm. This Gaussian spectral filter was not only of great help to reshape the pulse spectrum into initial state, but also important for maximizing the pulse quality and peak power[17]. Its central wavelength could be tuned by adjusting the reflective angle of the grating.

The design of Arm 2 was the same as Arm 1. The Mamyshev mode-locked mechanism could be realized by seperating the central wavelengths of the two spectral filters. In order to achieve high-peak-power laser pulses, low mode-locking threshold, and sufficient spectral broadening, the PBS2 would be served as the main output port of the laser, while the output power from PBS1 would be as week as possible. All passive fibres were the same type (HI1060), and hence all fibres used in this laser were single-mode and non-polarization-maintaining. Most of them were coiled into circles with the radius of more than 10 cm without active cooling condition. The fibre length was measured to be 7.2 m and the free-space optical length was about 1.28 m. Thus, the total optical length of this cavity was about 11.86 m.

The experimental results were measured by an oscilloscope (Tektronix, DPO5204B) with a photodetector (Thorlabs, DET08CL), a fibre-coupled power meter (THORLABS, PM20), a free-space power meter (Newport, 843-R-USB), an optical spectrum analyzer (Anritsu, MS9740A), a RF spectrum analyzer (Tektronix, RSA306B), and an autocorrelator (APE, pulseCheck USB). The output laser pulses were dechirped externally by a homemade compression system with a single blazed grating.

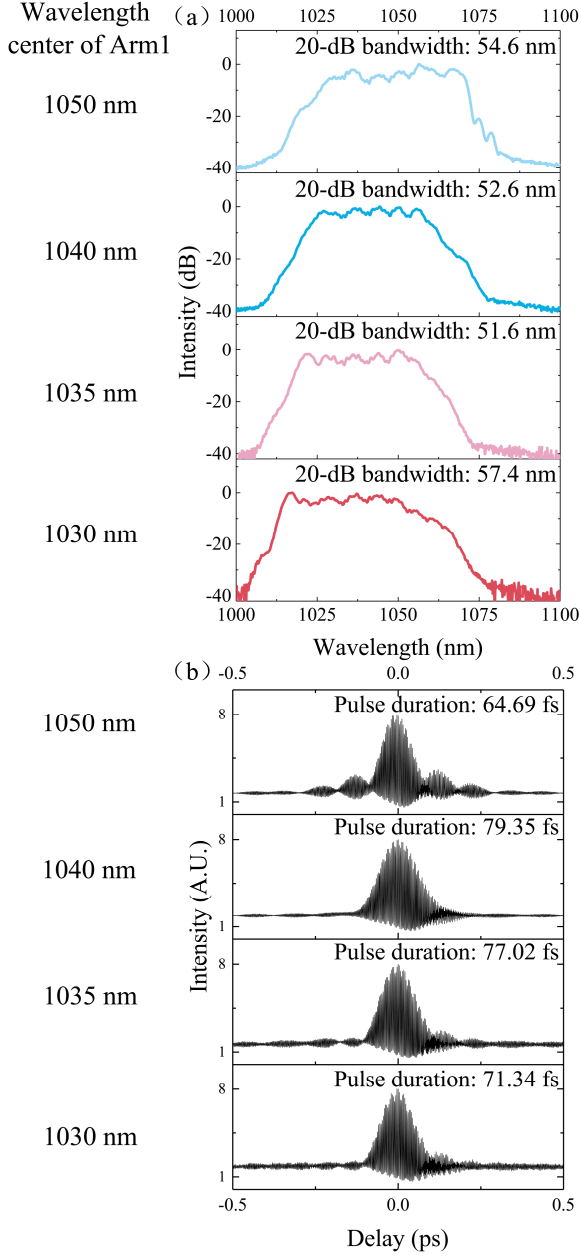

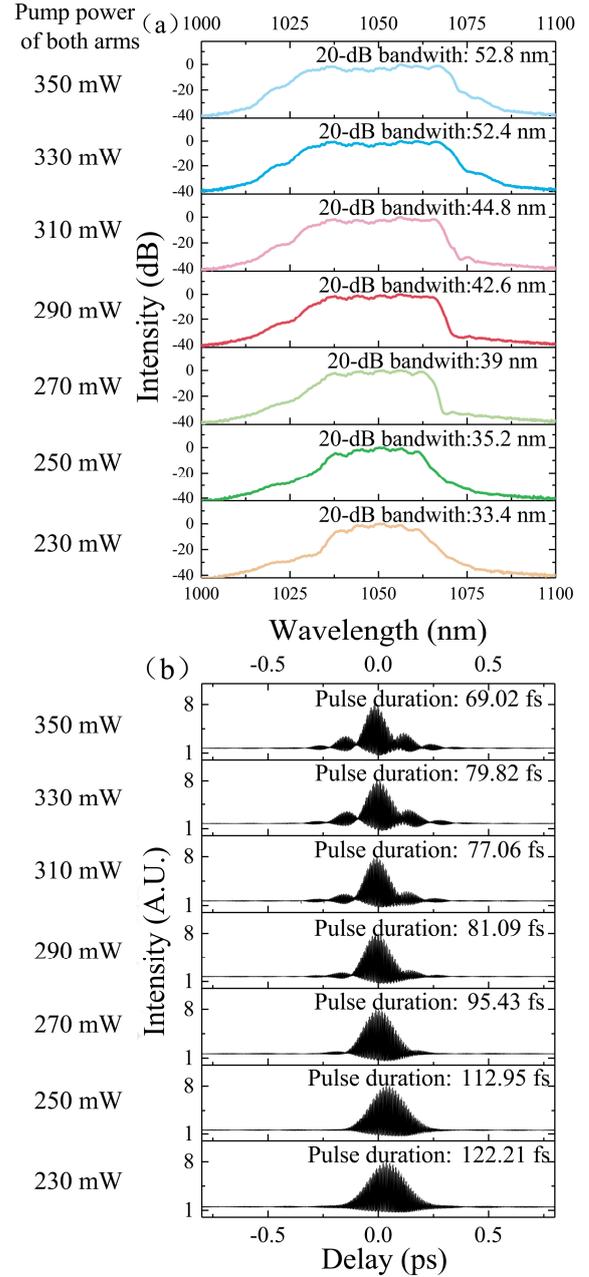

**Figure 2.** The characteristics of output pulses with different spetral-filter central wavelengths: (a) optical spectra; (b) interferometric autocorrelation traces after compression.

## 3. Results

In order to optimize the Mamyshev oscillator and obtain high-energy ultra-short pulses, the influences of spectral-filter central wavelength and pump power on the output laser characteristics have been investigated in our experiments.

### 3.1 Central wavelength of the spectral filters

First, we investigated the influences of the spectral filters. By adjusting the reflected angle of gratings, the central wavelengths of both filters could be tuned to a specified wavelength. In our experiments, the central wavelengths in Arm 1 were chosen to be 1030 nm, 1035 nm, 1040 nm, and 1050 nm, respectively. When the pump power was low, the overlap of two spectral filters was needed to start the laser operation. By rotating the QWP and HWP before the isolator in both arms, the NPE mechanism and hence initial mode-locked pulses could be achieved. While maintaining the mode-locked operation, the pump power of both arms could be increased gradually, and central wavelengths of two spectral filters were separated synchronously by changing the reflective angle of grating in Arm 2. Finally, another mode-locking state with a sudden broadening spectrum was obtained, which is based on the Mamyshev mechanism.

**Figure 3.** (a) Output spectra from PBS 2 and (b) interferometric autocorrelation traces of the dechirped pulses.

In this state, the optical spectra and interferometric autocorrelation traces of the output pulses with maximum output power are shown in Figure 2. The slight asymmetry of autocorrelation traces might be related to the inconsistency of two optical paths in the autocorrelator. In the cases of 1030 nm, 1035 nm, and 1040 nm, the laser could maintain in mode-locking operation when the pump power of both arms were increased to about 250 mW. With further increasing the pump power or separating the spectral filters, the laser operation would turn into harmonic mode-locking state. In these three cases, the laser components close to the peak of small signal gain spectrum (about 1030 nm) in the Yb-doped fibre can obtain higher gain. This resulted in a broadened output spectrum around 1030 nm, including lasing wavelength below 1025 nm where the reabsorption effect is serious. Figure 2(b) shows that these three interferometric autocorrelation traces and their calculated pulse duration after compression were similar; all the corresponding time-bandwidth product (TBP) were calculated to be about 0.7.

By tuning the centers of two spectral filters to longer wavelength, the Mamyshev oscillator could operate in stable mode-locking state at higher pump power of about 350 mW, as shown in the case of 1050 nm in Figure 2(a). In this case, the gain window shifted to the red side due to gain saturation and reabsorption effects in Yb-doped fibre. As a result, its lasing spectrum broadened at longer wavelength near 1050 nm, which was different from those three cases mentioned above. Figure 2(b) showed that the dechirped interferometric autocorrelation trace had more side-lobes. It indicates that the output pulses might contain more high-order dispersion components and accumulated nonlinear phase which were difficult to be compensated by a single grating. The dechirped laser pulses had a narrower pulse duration of 64.69 fs in this case. According to the spectral width and pulse duration after compression, the corresponding TBP was calculated to be 0.61. Both interferometric autocorrelation trace in Figure 2(b) and TBP showed that the pulses are well dechirped.

According to the output parameters, pulse evolution in the case of 1050 nm was close to the limit of similariton pulses, which was different from other three cases. We assume that these phenomena may be caused by the limited pump power and the presence of nonlinearity and gain-shaping. In our cavity when the pump power was limited and hence the total gain was low, cases with shorter spectral center could not reach the gain-managed nonlinearity (GMN) regime[24]. Besides, since the small-signal gain was high, their pulse evolution would be easily limited by the nonlinearity and gain-shaping effect [25]. In contrast, in the case with central wavelength far away from the peak of gain spectrum, pulse evolution would just distort slightly with higher pump power. Therefore, higher pulse energy and shorter pulse duration can be achieved in this case.

*3.2 Pump power*

We have also investigated the influences of pump power on the output performances of Mamyshev oscillator. In this experiment, the central wavelength of both spectral filters were tuned to 1050 nm. By adjusting the wave plates and output coupling ratios, the initial mode-locked pulses could be achieved with NPE mechanism. Subsequently, by separating the central wavelength of spectral filter in Arm 2 to 1044.4 nm and increasing the pump power of both arms to 260 mW gradually, another continuous-wave mode-locking state based on the Mamyshev mechanism was obtained. The 3-dB bandwidths of the spectral filters in Arm 1 and Arm 2 were measured to be 3.29 nm and 2.77 nm, respectively. This mode-locking state could be maintained when the pump power of both arms was increased to 350 mW. Multi-pulse operation was observed with further increasing the pump power, and the mode-locking state would be lost when the pump was decreased to 220 mW.

In this case, the settings of wave plates and spectral filters were kept after the Mamyshev mode-locking state was achieved. As a result, this oscillator could realize self-starting mode-locking as long as the pump power of both arms reached 260 mW. Two main reasons might lead to this self-starting operation. Firstly, the appropriate setting of optical elements in cavity could lead to the polarization conditions for Q-switched mode-locking based on NPE mechanism when the pump power was high enough. Secondly, the separation between the spectral center of two filters was not too large. Therefore, the intense pulse spikes could introduce enough spectral broadening to start the mode locking operation based on Mamyshev mechanism.

Figure 3(a) and 3(b) show the output spectra and interferometric autocorrelation traces of the dechirped pulses under different pump power, respectively. With the increase of pump power, the spectrum experienced significantly broadening, while the pulse duration decreased gradually. In this process, the side-lobes of autocorrelation trace became obvious, which were resulted from the accumulation of high-order dispersion and nonlinear phase.

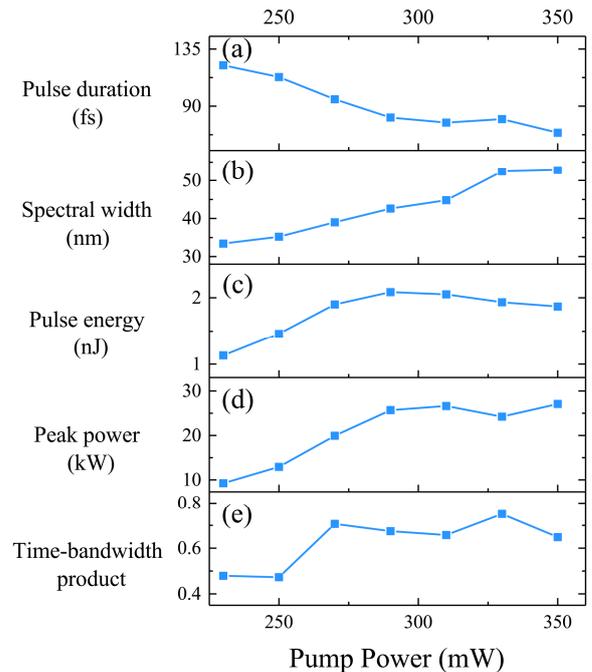

**Figure 4.** Output pulse properties at different pump power: (a) dechirped pulse duration; (b) spectral width; (c) single pulse energy; (d) dechirped pulse peak power; (e) dechirped pulse time-bandwidth product.

Figure 4(a)-(e) summarize the trend of pulse duration, spectral width, single pulse energy, peak power and TBP with different pump. With the increase of pump power, the single pulse energy increased linearly at first, and then slightly decreased when the pulse energy was up to 2.08 nJ. We speculate that the decrease of pulse energy may result from the increase of laser peak power. The laser with higher peak power led to significant effects on the nonlinear phase shift, which decreased the transmittance of isolator in the cavity and hence worked against pulse formation[25]. In addition, the peak power of the dechirped pulse increased gradually with the increase of pump power, except for a slight decrease at 330 mW. This might be caused by the larger pulse duration in the case of 330 mW where high-order dispersion components could not be compensated perfectly. This conjecture was also supported by the TBP shown in Figure 4(e). Both Figure 3(b) and Figure 4(e) show that the output pulses were well dechirped.

In conclusion of this experiment, the laser pulses were gradually transformed from typical similaritons with linear chirp to the limit of similariton pulses but not GMN regime when pump power was increased. This has further confirmed the assumption in part A: the limited pump power and the presence of nonlinearity and gain-shaping had important influences on the pulse evolution of our laser. Moreover, in this case with the central wavelength far away from the peak of gain spectrum, the self-starting operation could be achieved with total pump power of only 520 mW. This phenomenan might benefit from the appropriate polarization conditions, small spectral filter separation and higher accepted pump power.

### 3.3 Optimized output results

Under the guidance of part A and B mentioned above, we have optimized the cavity parameters of Mamyshev oscillator in order to achieve higher output power and shorter pulsed duration. In this case, the central wavelength of two spectral filters were adjusted to be 1050 nm and 1044.4 nm, with their 3-dB bandwidth of 3.29 nm and of 2.77 nm, respectively. The output coupling ratio of arm 2 was designed to be 92.72%.

When the pump power reached 370 mW in Arm 1 and 360 mW in Arm 2, the laser could maintain in the continuous-wave mode-locking state. Typical pulse train is shown in Figure 5(a); those weak fluctuations might be induced by the limited sampling rate of our photodetector and oscilloscope. In order to further verify the short-term stability of the laser, an RF spectrum was measured and shown in Figure 5(b). The RF spectrum had a high signal-to-noise ratio of ~70 dB, which indicates high stability of the mode-locking state. No multi-pulse operation or harmonic mode-locking could be found in the long-span RF spectrum. Besides, the period of output pulses and the fundamental repetition rate of the laser were measured to be 41 ns and 24.39 MHz, respectively, which were consistent with the optical cavity length of 11.86 m.

The optical spectrum and autocorrelation trace of the output pulses are shown in Figure 6. The 20-dB bandwidth of the spectrum was measured to be 54.6 nm. The FWHM of the intensity autocorrelation trace before compression was measured to be 1.06 ps, corresponding to the pulse duration of 0.75 ps. The FWHM of dechirped pulse was calculated to be ~64.69 fs by assuming a Gaussian shape. Noticeable side-lobes could be found in the interferometric autocorrelation which might be caused by uncompensated high-order dispersion and accumulated nonlinear phase. The average output power from PBS2 was 46.55 mW, corresponding to a pulse energy of 1.89 nJ. 73% of the pulse energy was retained after the pulse compression, and hence dechirped pulses with their peak power of 21.3 kW were achieve.

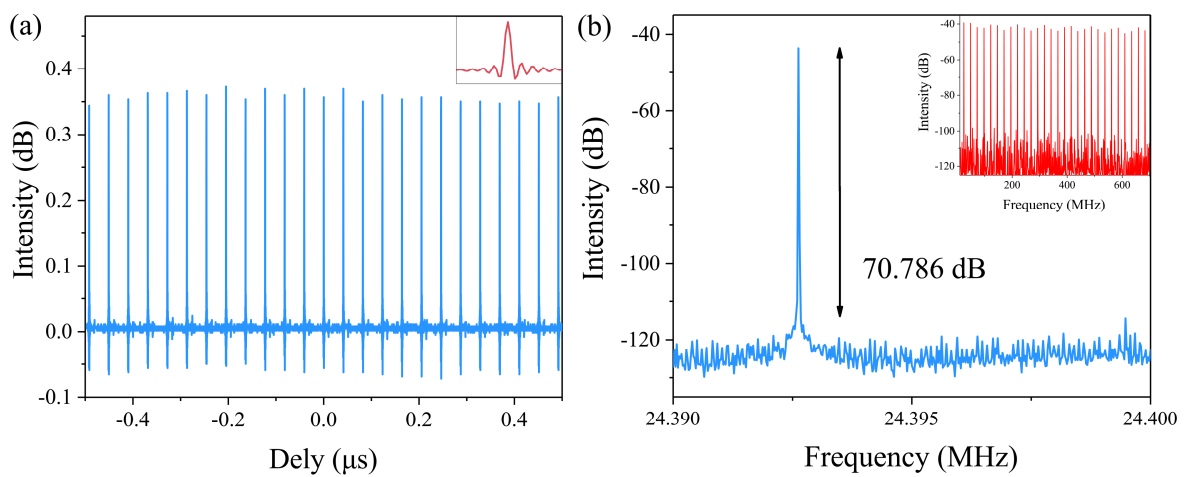

**Figure 5.** The output pulse properties: (a) mode-locked pulse train, inset: pulse trace in a cavity period; (b) RF spectrum with 1-Hz resolution bandwidth in a span range of 10 kHz, inset: RF spectrum with 100-Hz resolution bandwidth in a span range of 700 MHz.

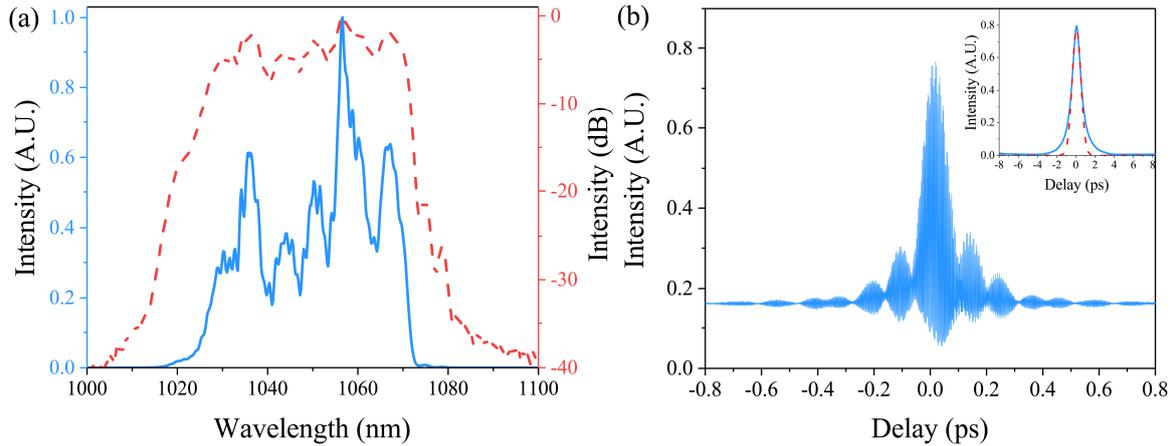

**Figure 6.** The output pulse properties: (a) output spectrum in linear (bule line) and logarithmic (red dots) scales; (b) interferometric autocorrelation trace of compressed pulses; inset: intensity autocorrelation trace of the output pulse (bule line) and the Gaussian fitting curve (red dots).

## 4. Conclusion

In conclusion, we have demonstrated a mode-locked Yb-doped Mamyshev oscillator with low threshold. We have investigated the influences of the spectral-filter central wavelength and the pump power on the laser output characteristics. When the central wavelength of spectral filters were far away from the peak of gain spectrum and the pump power was high, laser pulses would perform a transitional state between similariton evolution and GMN regime. In this case, benefit from the appropriate polarization conditions, small spectral filter separation and higher accepted pump power, the laser could realize the self-starting operation and remain continuous-wave mode-locking state within a wide range of pump power. With the optimization, the Mamyshev oscillator could produce laser pulses with 20-dB spectral width of 54.6 nm and pulse energy of 1.89 nJ. The output pulses have been dechirped externally to 64.69 fs with peak power of 21.3 kW. Our results have provided a new perspective for studying the pulse evolution and self-starting mechanism of Mamyshev oscillators with low threshold. However, NPE and hence non-polarizaiton-maintaining fibres is needed in this self-starting operation, which indicates the mode-locking state will be sensitive to the environmental perturbations. Future works will focus on the investigation of environmental-stable self-starting operation and different pulse evolution states in the Mamyshev oscillators.

## Acknowledgements

The authors would like to acknowledge Fujuan Wang and Jiaoyang Li for the technical support in this work. This work was partially supported by National Natural Science Foundation of China under grant No. 12074441 and 11774438 as well as Guangdong Basic and Applied Basic Foundation in China under grant No. 2019A1515011572.